\documentclass[pr,cha,onecolumn,superscriptaddress,preprintnumbers,showpacs, amsmath,amssymb]{revtex4-1}
\usepackage{bm}
\usepackage[colorlinks=true,linkcolor=blue,citecolor=blue]{hyperref}
\usepackage{amsmath}
\usepackage{amssymb}
\usepackage{amsthm}
\usepackage{amsfonts}
\usepackage{enumerate}
\usepackage{latexsym}
\usepackage{ifpdf}
\usepackage{graphicx}
\usepackage{makeidx}
\expandafter\ifx\csname package@font\endcsname\relax\else
\expandafter\expandafter
\expandafter\usepackage
\expandafter\expandafter
\expandafter{\csname package@font\endcsname}
\fi
\usepackage{color}
\usepackage{times}

\begin{document}

\title{Synthesis of single-crystalline LuN films}

\author{Guanhua Su}
\thanks{These authors contributed equally to this work}
\affiliation{Ningbo Institute of Materials Technology and Engineering, Chinese Academy of Sciences, Ningbo 315201, China}
\affiliation{Center of Materials Science and Optoelectronics Engineering, University of Chinese Academy of Sciences, Beijing 100049, China}
\author{Shuling Xiang}
\thanks{These authors contributed equally to this work}
\affiliation{Ningbo Institute of Materials Technology and Engineering, Chinese Academy of Sciences, Ningbo 315201, China}
\author{Jiachang Bi}
\affiliation{Ningbo Institute of Materials Technology and Engineering, Chinese Academy of Sciences, Ningbo 315201, China}
\affiliation{Center of Materials Science and Optoelectronics Engineering, University of Chinese Academy of Sciences, Beijing 100049, China}
\author{Fugang Qi}
\affiliation{Ningbo Institute of Materials Technology and Engineering, Chinese Academy of Sciences, Ningbo 315201, China}
\author{Peiyi Li}
\affiliation{Ningbo Institute of Materials Technology and Engineering, Chinese Academy of Sciences, Ningbo 315201, China}
\author{Shunda Zhang}
\affiliation{Ningbo Institute of Materials Technology and Engineering, Chinese Academy of Sciences, Ningbo 315201, China}
\affiliation{Center of Materials Science and Optoelectronics Engineering, University of Chinese Academy of Sciences, Beijing 100049, China}
\author{Shaozhu Xiao}
\affiliation{Ningbo Institute of Materials Technology and Engineering, Chinese Academy of Sciences, Ningbo 315201, China}
\author{Ruyi Zhang}
\affiliation{Ningbo Institute of Materials Technology and Engineering, Chinese Academy of Sciences, Ningbo 315201, China}
\author{Zhiyang Wei}
\affiliation{Ningbo Institute of Materials Technology and Engineering, Chinese Academy of Sciences, Ningbo 315201, China}
\affiliation{Center of Materials Science and Optoelectronics Engineering, University of Chinese Academy of Sciences, Beijing 100049, China}
\author{Yanwei Cao}
\email{ywcao@nimte.ac.cn}
\affiliation{Ningbo Institute of Materials Technology and Engineering, Chinese Academy of Sciences, Ningbo 315201, China}
\affiliation{Center of Materials Science and Optoelectronics Engineering, University of Chinese Academy of Sciences, Beijing 100049, China}

\date{\today}

\begin{abstract}

In the nitrogen-doped lutetium hydride (Lu-H-N) system, the presence of Lu-N chemical bonds plays a key role in the emergence of possible room-temperature superconductivity at near ambient pressure. However, due to the synthesis of single-crystalline LuN being a big challenge, the understanding of LuN is insufficient thus far. Here, we report on the epitaxial growth of single-crystalline LuN films. The crystal structures of LuN films were characterized by high-resolution X-ray diffraction. The measurement of low-temperature electrical transport indicates the LuN film is semiconducting from 300 to 2 K, yielding an activation gap of $\sim$ 0.02 eV. Interestingly, negative magnetoresistances can be observed below 12 K, which can result from the defects and magnetic impurities in LuN films. Our results uncover the electronic and magnetic properties of single-crystalline LuN films.

\end{abstract}

\maketitle
\newpage

\section{Introduction}

Very recently, room-temperature superconductivity-like transitions at the near ambient pressure were observed in a mixture of LuH$_{3-\delta}$N$_\varepsilon$ (92.25\% purity) and LuN$_{1-\delta}$H$_\varepsilon$ (7.29\% purity) compounds which have both N-substitution and H-vacancy defects \cite{Nature-2023-Dias}. This amazing report immediately ignites the worldwide interest of studying Lu-H-N compounds \cite{Nature-2023-Dias,arXiv-2023-Jin,arXiv-2023-LiuP, CPL-2023-Cheng, arXiv-2023-Wen,arXiv-2023-Sun,arXiv-2023-Liu, arXiv-2023-Cheng,arXiv-2023-Wen2,SB-2023-Cheng,arXiv-2023-Boeri,AIP-2023-Zhang,arXiv-2023-MLiu,arXiv-2023-Ho,arXiv-2023-Cui,arXiv-2023-Zurek,arXiv-2023-MM,arXiv-2023-Heil,ArXiv-2020-Duan,Research-2022-Cui,IC-2021-Cui,M-2021-Wang,arXiv-2023-LuT,arXiv-2023-Monserrat,arXiv-2023-LiP,arXiv-2023-Tao,arXiv-2023-Wen3,arXiv-2023-Errea,arXiv-2023-Fiory,arXiv-2023-Hemley,arXiv-2023-Guo,arXiv-2023-Salke,arXiv-2023-Gubler,arXiv-2023-Pav,arXiv-2023-Peng}.  Interestingly, similar high-temperature transitions (with non-zero resistance) can be repeated independently by other researchers \cite{arXiv-2023-Peng}. As highlighted, the presence of Lu-N chemical bonds plays a key role in stabilizing the crystal structure and inducing possible room-temperature superconductivity at the near ambient pressure \cite{Nature-2023-Dias,arXiv-2023-Ho,arXiv-2023-Pav}. However, most experimental studies of the Lu-H-N system focus on Lu-H not Lu-N compounds at present.

Generally, bulk LuN with a rock-salt crystal structure can only be synthesized at high temperatures and high pressures (such as 2000 K and 30 GPa) in the diamond anvil cell \cite{JAC-2009-Niwa}. As the last and heaviest element of the lanthanide rare earth family, the electronic configuration of Lu is 4$f^{14}$5$d^{1}$6$s^{2}$. Therefore, Lu$^{3+}$ has a fully filled 4$f$ shell, indicating there is no net magnetic moment in ideal LuN. More interestingly, in rare-earth (RE) nitrides from 57 (La) to 71 (Lu), the $f$ electrons become more localized with increasing atomic numbers. The Coulomb interaction U can be as large as 10.9 eV in LuN \cite{PRB-2007-Larson}. At ambient pressure, the stable phase of bulk LuN has a rock-salt crystal structure with the cubic $ Fm\bar{3}m $ space group \cite{JPCS-2015-Singh}. The lattice parameter of bulk LuN is $a$ $\sim$ 4.76 \AA, which does not match common oxide substrates with a lattice parameter $a$ $\sim$ 4.0 \AA for epitaxial growth \cite{JCG-2017-Uec}. Due to the synthesis of single-crystalline LuN being a big challenge, the understanding of its properties is insufficient thus far \cite{JAC-2009-Niwa,JPCS-2015-Singh,PRB-2007-Larson,JMR-2004-Sue,AIP-2022-Devese,PRB-2009-Granville}.

\begin{figure*}[t]
\includegraphics[width=0.95\textwidth]{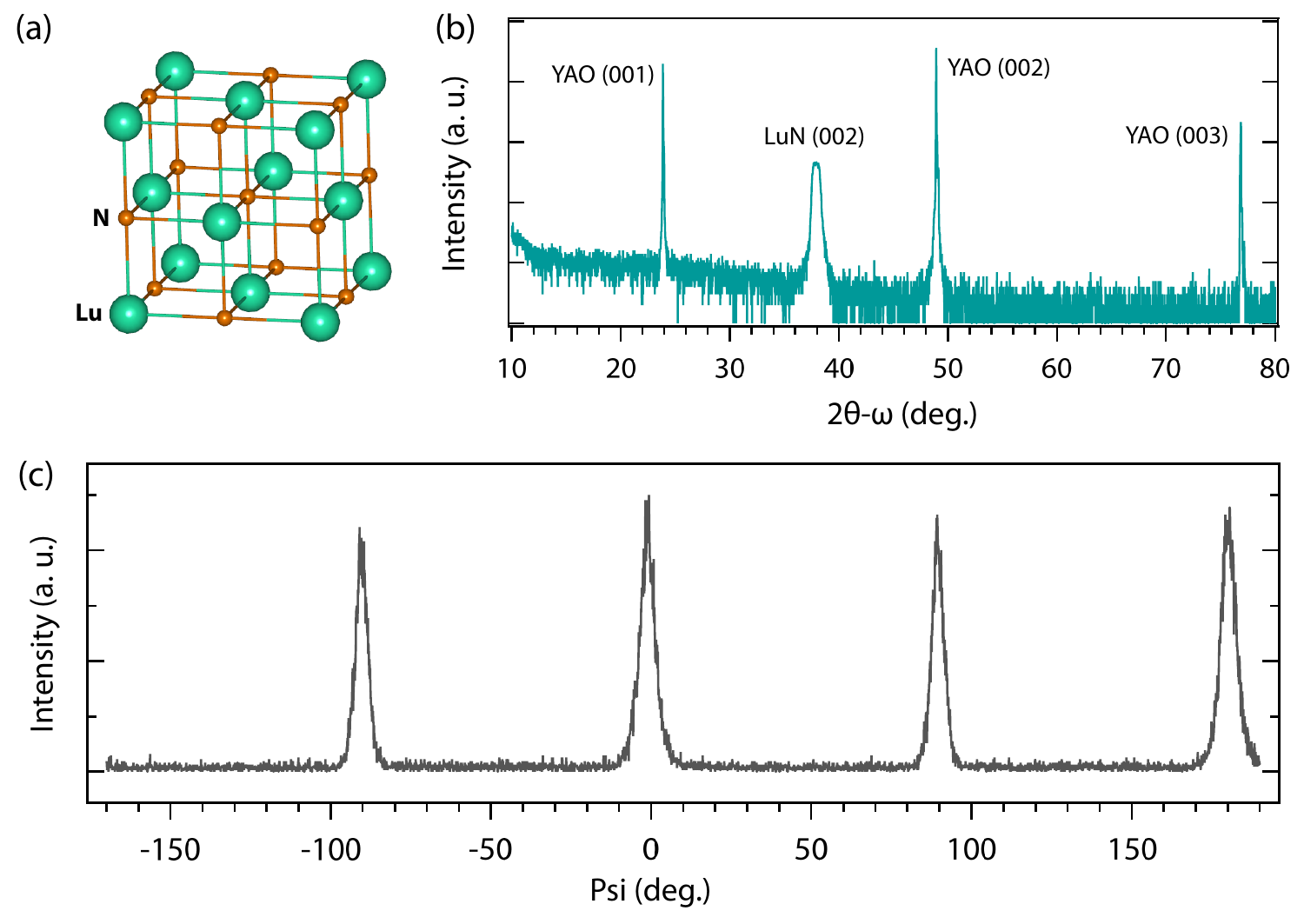}
\caption{\label{} (a) Schematic of cubic LuN crystal structure. (b)  Wide-range 2$\theta$-$\omega$ scan of the LuN film on the YAO substrate. (c) Psi scan of the LuN film on the YAO substrate.}
\end{figure*}

\begin{figure*}[]
\includegraphics[width=0.5\textwidth]{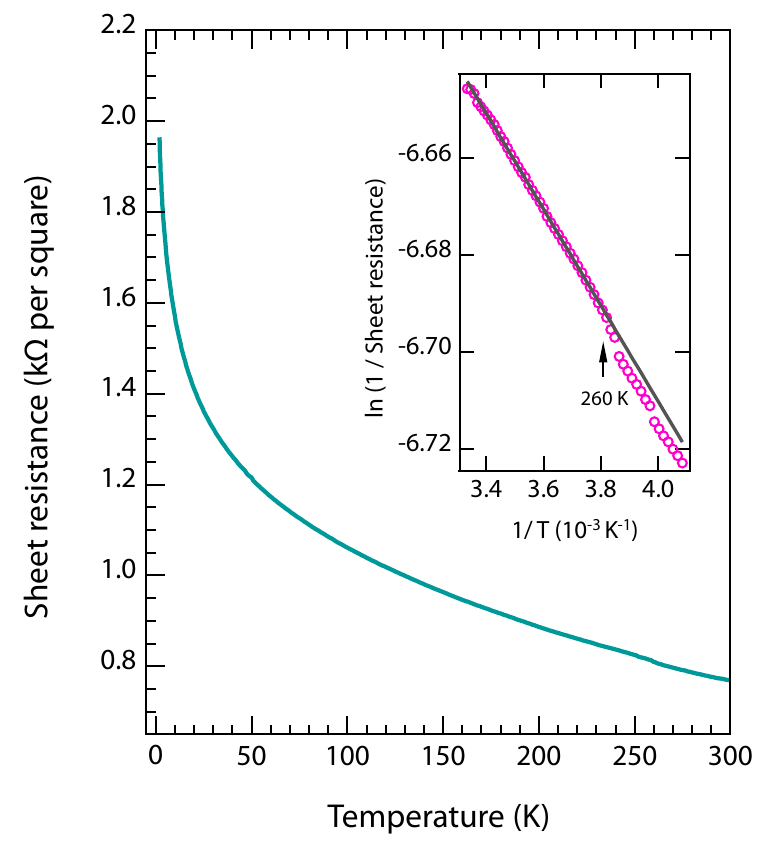}
\caption{\label{} Temperature-dependent sheet resistance of the LuN film from 300 to 2 K.}
\end{figure*}

To address the above, we synthesized single-crystalline LuN films by magnetron sputtering. The measurement of high-resolution X-ray diffraction (XRD) indicates the cubic crystal structure of LuN films. The characterization of low-temperature electrical transport reveals the LuN film is semiconducting, which is consistent with the calculated value by the first-principle calculations. Interestingly, negative magnetoresistances can be observed below 12 K, which can result from defects or magnetic impurities. Our results uncover the electronic and magnetic properties of single-crystalline LuN films.

\section{Experiments and calculations}

Single-crystalline LuN films (thickness $\sim$ 230 nm) were epitaxially grown on (110)-oriented YAlO$_3$ (YAO) single-crystal substrates (5$\times$5 $\times$0.5 mm$^{3}$) by a home-made radio frequency (RF) magnetron sputtering, the setup of which is analogous to our previous reports \cite{PRM-2021-Bi,APLM-2021-Zhang,NL-2023-Zhang}. The base vacuum pressure of the sputtering system is better than 1 $\times$ 10$^{-7}$ Torr, and the purities of the 2-inch Lu target and the reactive gas are 99.99\% and 99.999\%, respectively. During the growth process, the substrate was held at 950 $^{\circ}$C. The pressure of the  reactive gas with a mixture of Ar and N$_2$ (9:1 ratio)  was kept at 20 mTor with a flow rate of 5.4 sccm. The crystal structures of LuN films were characterized by high-resolution XRD (Bruker D8 Discovery) with the Cu K$\alpha$ source ($\lambda$ = 1.5405 \AA). The electrical transport properties of LuN films were measured from 300 to 2 K by Physical Property Measurement System (PPMS) in a van der Pauw geometry (DynaCool, Quantum Design).

Our first-principle calculations were accomplished based on density-functional theory (DFT), using the VASP package. Perdew-Burke-Ernzerhof (PBE), one kind of generalized gradient approximation (GGA) was applied to describe the electron-ion interactions. A primitive cell including one Lu and one N was adopted in the calculations. K-points over the Brillouin zone were 18 $\times$ 18 $\times$ 18 and 27 $\times$ 27 $\times$ 27 for structure optimization and static self-consistent computation, ensuring calculation accuracy. The convergence criteria of total energy and force components were set to be 1 $\times$ 10$^{-6}$ eV and 0.01 eV/\AA, respectively, whereas 420 eV was selected as the cut-off energy of the plane-wave basis. For this strongly correlated electron system, we did a bandgap correction by applying the PBE+U scheme.

\begin{figure*}[]
\includegraphics[width=0.5\textwidth]{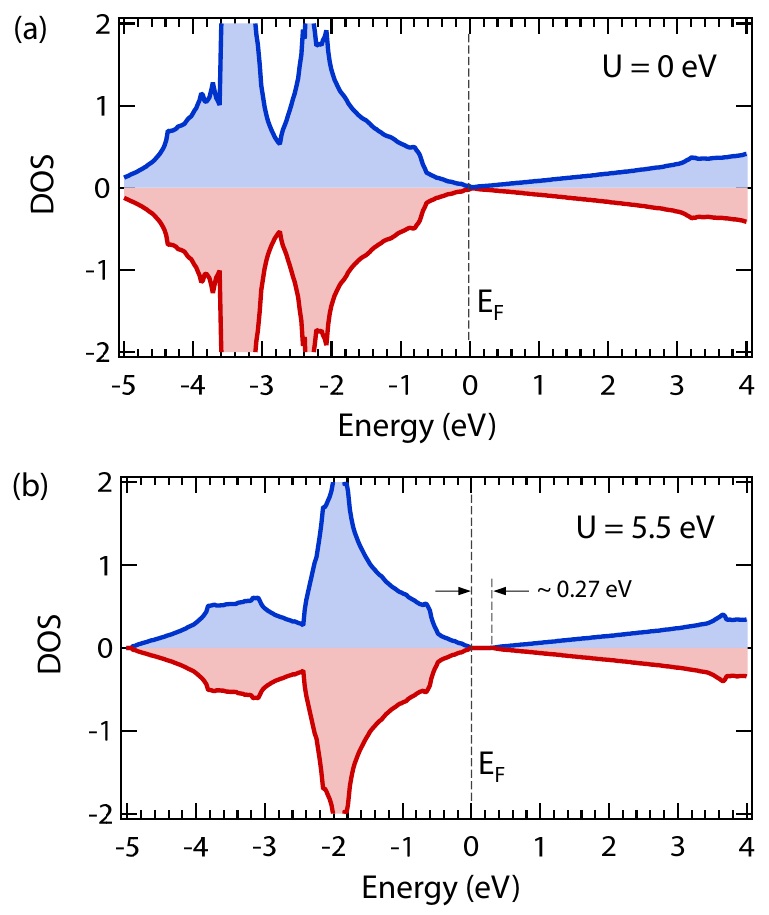}
\caption{\label{} Calculated density of states of LuN without (a) and with (b) Hubbard interaction U.}
\end{figure*}

\begin{figure*}[th]
	\includegraphics[width=0.95\textwidth]{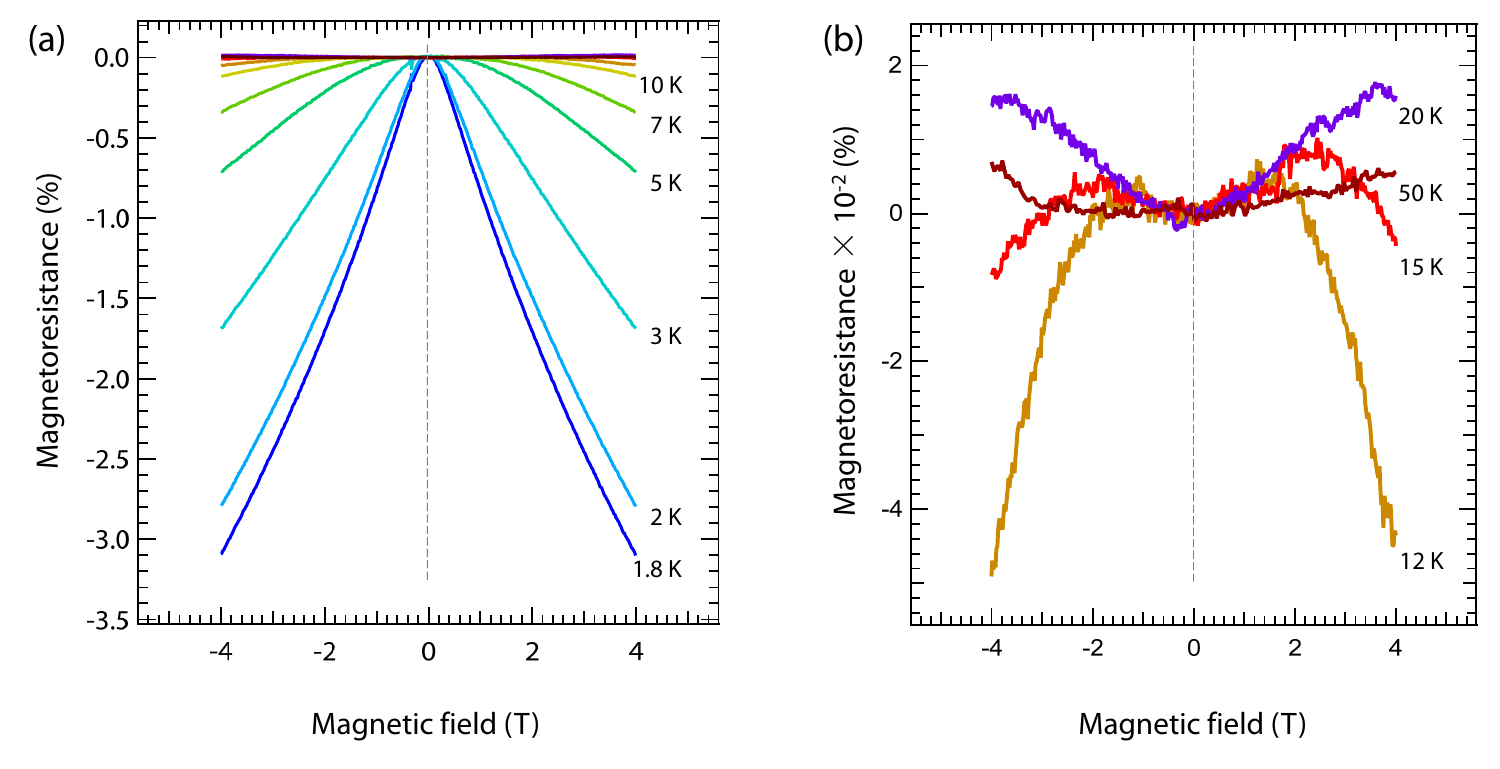}
	\caption{\label{} Temperature-dependence of magnetoresistances of LuN films. (a) 1.8 - 10 K. (b) 12 - 50 K.}
\end{figure*}
\section{Results and Discussion}

First, we investigate the crystal structures of LuN films by high-resolution XRD. As seen in Fig. 1(b), only one film peak (002) near 37.905$^{\circ}$ can be observed in the wide-range 2$\theta$-$\omega$ scan of LuN films on YAO substrates, corresponding to a lattice constant $c$ $\sim$ 4.74 \AA, which is consistent with the lattice constant of bulk LuN ($\sim$ 4.76 \AA) \cite{JPCS-2015-Singh}. It is noteworthy that bulk YAO has an orthorhombic structure with lattice parameters $a_{o}$ = 5.176 \AA, $b_{o}$ = 5.307 \AA, and $c_{o}$ = 7.355 \AA. Therefore, in view of pseudocubic symmetry, the lattice constants of the YAO (110) plane are $a_{ps}$ = 3.678 \AA~ and $b_{ps}$ = 3.707 \AA. Since the lattice parameter of bulk LuN is near 4.76 \AA, the epitaxial growth of LuN (001) films is available with a 45$^{\circ}$ rotation of crystal structure on YAO (110) substrates. To confirm the symmetry of LuN films, we performed a Psi scan of the LuN (111) crystal plane (see Fig. 1(c)), indicating the four-fold symmetry of LuN films, which further demonstrates the cubic symmetry of LuN films.

Next, we characterize the electronic properties of LuN films by measuring low-temperature electrical transport. As seen in Fig. 2, with decreasing the temperature from 300 to 1.8 K, the sheet resistance of the LuN film increases from 0.77 k$\Omega$/$\Box$ at 300 K to 1.96 k$\Omega$/$\Box$ at 1.8 K, indicating a semiconducting behavior. The estimated activation gap by fitting the conductance (see the inset in Fig. 2) is $\sim$ 0.02 eV. To further understand the electronic structure of LuN, we performed the first-principle calculations. As shown in Fig. 3, the Coulomb interaction U plays a vital role in the electronic structure of LuN. Theoretically, LuN is metallic when U = 0, whereas there is an opened gap with increasing U. The calculated band gap is 0.27 eV when applying U$_f$ = 5 .5 eV to Lu-$f$ orbitals, which is consistent with previously calculated indirect band gaps (such as 0.14 and 0.36 eV) \cite{JPCS-2015-Singh,PRB-2007-Larson}. The deviation between experiments and calculations can be due to the presence of defects and impurities in real materials.
In addition, it is noted that the DOS of spin-up and spin-down are perfectly symmetrical due to the fully filled 4$f$ shell of Lu$^{3+}$.

At last, we probe the magnetic properties of LuN films by carrying out measurements of temperature-dependent magnetoresistances. Here, the magnetoresistance is defined as $R(B) - R(B=0) / R(B=0)$ with applying the magnetic field (perpendicular to the sample surface) from -4 T to 4 T. Unexpectedly,  as shown in Fig. 4, a distinct negative magnetoresistance can be observed at the temperature 1.8 to 12 K, which is a contrast to the physical picture of fully filled 4$f$-shell of Lu$^{3+}$ in ideal LuN (see Fig. 3). Interestingly, with increasing the temperature from 1.8 to 12 K, the signal of magnetoresistance is strongly suppressed. Near 15 K, the sign of magnetoresistance becomes very weak, and changes from negative to positive. There are several origins leading to the emergence of negative magnetoresistance behavior in nonmagnetic materials: (a) chiral anomaly in Weyl semimetals, (b) current jetting effects, (c) weak localization effect, and (d) ferromagnetic impurities in samples \cite{NC-2016-Li,NC-2016-Arnold,PRL-2007-Gorbachev,PRB-2004-May,SR-2021-Pal}. Here, on the one hand, the signal of negative magnetoresistances is very small ($\sim$ 2\%). On the other hand, there are several magnetic purities (such as Ni and Fe) at the level of 30 ppm in the Lu metal, which can lead to significant spin-glass transitions near 200 K \cite{AIP-2023-Zhang}. Therefore, the observation of negative magnetoresistances at low temperatures in LuN films can arise from magnetic impurities and defects in LuN films.

\section{Conclusion}

In summary, we successfully synthesized single-crystalline LuN films on YAO substrates by magnetron sputtering. The crystal and electronic strcutures of LuN films were investigated by high-resolution XRD, low-temperature electrical transport, and the first principle calculations. In this work, the LuN film is semiconducting with an activation gap $\sim$ 0.02 eV. More interestingly, distinct negative magnetoresistances can be observed below 12 K, resulting from defects or magnetic impurities. Our results uncover electronic and magnetic properties of single-crystalline LuN films.

\section{Acknowledgments}

We acknowledge insightful discussions with Jiandong Guo, Jiandi Zhang, Er-jia Guo, Bing Shen, Xiong Yao, and Rui Peng. This work was supported by the National Key R\&D Program of China (Grant No. 2022YFA1403000), the National Natural Science Foundation of China (Grant Nos. U2032126, 11874058, and U2032207), the Pioneer Hundred Talents Program of the Chinese Academy of Sciences, the Zhejiang Provincial Natural Science Foundation of China under Grant No. LXR22E020001, the Beijing National Laboratory for Condensed Matter Physics, the Ningbo Natural Science Foundation (Grant No. 2022J292), and the Ningbo Science and Technology Bureau (Grant No. 2022Z086).

\newpage

\end{document}